\documentclass[final,5p,times,twocolumn]{elsarticle}

\biboptions{sort&compress}

\usepackage{amsmath}
\usepackage{amssymb}
\usepackage{booktabs}
\usepackage{graphicx}
\usepackage{hyperref}
\usepackage{xcolor}
\usepackage{microtype}
\usepackage{array}

\hypersetup{
  colorlinks=true,
  linkcolor=blue!70!black,
  citecolor=blue!70!black,
  urlcolor=blue!70!black
}

\journal{Computer Physics Communications}

\providecommand{\FleetPeakVsFFTWmin}{}\renewcommand{\FleetPeakVsFFTWmin}{1.2}
\providecommand{\FleetPeakVsFFTWmax}{}\renewcommand{\FleetPeakVsFFTWmax}{2.8}
\providecommand{\FleetHierVsFFTWmin}{}\renewcommand{\FleetHierVsFFTWmin}{20}
\providecommand{\FleetHierVsFFTWmax}{}\renewcommand{\FleetHierVsFFTWmax}{33}
\providecommand{\FleetHierVsPeakmin}{}\renewcommand{\FleetHierVsPeakmin}{12}
\providecommand{\FleetHierVsPeakmax}{}\renewcommand{\FleetHierVsPeakmax}{18}

\providecommand{\ZenFleetFFTW}{}\renewcommand{\ZenFleetFFTW}{32.84}
\providecommand{\ZenFleetPeak}{}\renewcommand{\ZenFleetPeak}{22.78}
\providecommand{\ZenFleetHier}{}\renewcommand{\ZenFleetHier}{1.24}
\providecommand{\ZenFleetPeakVsFFTW}{}\renewcommand{\ZenFleetPeakVsFFTW}{1.4}
\providecommand{\ZenFleetHierVsFFTW}{}\renewcommand{\ZenFleetHierVsFFTW}{26}

\providecommand{\ZenScalePeakVsAoclmin}{}\renewcommand{\ZenScalePeakVsAoclmin}{1.2}
\providecommand{\ZenScalePeakVsAoclmax}{}\renewcommand{\ZenScalePeakVsAoclmax}{4.1}

\providecommand{\ZenScalePeakA}{}\renewcommand{\ZenScalePeakA}{0.34}

\providecommand{\GpuRoofCopy}{}\renewcommand{\GpuRoofCopy}{198}

\providecommand{\GpuRoofUnfusedBWmin}{}\renewcommand{\GpuRoofUnfusedBWmin}{159}
\providecommand{\GpuRoofUnfusedBWmax}{}\renewcommand{\GpuRoofUnfusedBWmax}{270}
\providecommand{\GpuRoofPeakVsUnfusedmin}{}\renewcommand{\GpuRoofPeakVsUnfusedmin}{6}
\providecommand{\GpuRoofPeakVsUnfusedmax}{}\renewcommand{\GpuRoofPeakVsUnfusedmax}{13}
\providecommand{\GpuRoofPeakVsRocmin}{}\renewcommand{\GpuRoofPeakVsRocmin}{3.6}
\providecommand{\GpuRoofPeakVsRocmax}{}\renewcommand{\GpuRoofPeakVsRocmax}{6.2}
\providecommand{\GpuRoofHierVsRocmin}{}\renewcommand{\GpuRoofHierVsRocmin}{7}
\providecommand{\GpuRoofHierVsRocmax}{}\renewcommand{\GpuRoofHierVsRocmax}{43}

\providecommand{\GpuRoofNfourkRoc}{}\renewcommand{\GpuRoofNfourkRoc}{3.00}
\providecommand{\GpuRoofNfourkFull}{}\renewcommand{\GpuRoofNfourkFull}{2.65}
\providecommand{\GpuRoofNfourkPeak}{}\renewcommand{\GpuRoofNfourkPeak}{0.48}
\providecommand{\GpuRoofNfourkHier}{}\renewcommand{\GpuRoofNfourkHier}{0.14}
\providecommand{\GpuRoofNfourkPeakVsRoc}{}\renewcommand{\GpuRoofNfourkPeakVsRoc}{6.2}
\providecommand{\GpuRoofNfourkHierVsRoc}{}\renewcommand{\GpuRoofNfourkHierVsRoc}{21}
\providecommand{\MtwoNfourkPeakVsMlx}{}\renewcommand{\MtwoNfourkPeakVsMlx}{3.1}
\providecommand{\MtwoNfourkHierVsMlx}{}\renewcommand{\MtwoNfourkHierVsMlx}{19}
\providecommand{\MtwoHierVsMlxmin}{}\renewcommand{\MtwoHierVsMlxmin}{8}
\providecommand{\MtwoHierVsMlxmax}{}\renewcommand{\MtwoHierVsMlxmax}{29}

\providecommand{\AccMedianFirst}{}\renewcommand{\AccMedianFirst}{4.26 \times 10^{-8}}
\providecommand{\AccMedianLast}{}\renewcommand{\AccMedianLast}{7.10 \times 10^{-8}}
\providecommand{\AccNfirst}{}\renewcommand{\AccNfirst}{1024}
\providecommand{\AccNlast}{}\renewcommand{\AccNlast}{65536}
\providecommand{\AccMax}{}\renewcommand{\AccMax}{2.30 \times 10^{-7}}

\providecommand{\AccTrials}{}\renewcommand{\AccTrials}{640}

\providecommand{\FleetThrLo}{}\renewcommand{\FleetThrLo}{4}
\providecommand{\FleetThrHi}{}\renewcommand{\FleetThrHi}{6.5}
\providecommand{\FleetThrLoMin}{}\renewcommand{\FleetThrLoMin}{1.0}
\providecommand{\FleetThrLoMax}{}\renewcommand{\FleetThrLoMax}{1.3}
\providecommand{\FleetThrHiMin}{}\renewcommand{\FleetThrHiMin}{28}
\providecommand{\FleetThrHiMax}{}\renewcommand{\FleetThrHiMax}{83}

\begin{document}

\begin{frontmatter}

\title{Peak Performance in Matched Filtering: Fused Correlation and Statistically Controlled Screening on CPUs and GPUs}

\author[syracuse]{Alexander H. Nitz\corref{cor1}}
\ead{ahnitz@syr.edu}
\author[syracuse]{Keisi Kacanja}
\author[syracuse]{Collin D. Capano}
\cortext[cor1]{Corresponding author.}

\address[syracuse]{Department of Physics, Syracuse University, Syracuse, NY 13244, USA}

\begin{abstract}
Batched matched filtering and cross-correlation over large template banks are a primary computational cost in gravitational-wave astronomy, observational seismology, and radio transient searches. Standard pipelines pair a general-purpose inverse Fast Fourier Transform (FFT) with separate array kernels for frequency-domain multiplication and peak finding, writing intermediate spectra and full correlation time series to memory for every template. We fuse the conjugate product into the first inverse-transform butterflies and the peak reduction into the last so that only candidate peaks leave the processor registers. We further tile batches in two dimensions so that $K^2$ correlations reuse $2K$ spectra in cache, shifting full-bandwidth matched filtering from a DRAM-bandwidth-bound regime to an on-chip cache-bandwidth-bound regime. In some situations, the signal-to-noise ratio (SNR) is concentrated toward lower frequencies while higher frequencies dominate the transform length. We develop a hierarchical screening method that takes advantage of this using a sequence of short low-frequency transforms before evaluating the full bandwidth. Per-stage screening thresholds are derived from the expected spectral power profile and matched-filter noise autocovariance to target a specified false-dismissal probability. We demonstrate these methods in the open-source \texttt{matchedfilter} library across eight CPU architectures and two integrated GPUs (AMD Radeon 8060S via Vulkan and Apple M2 via Metal). On an AMD Ryzen AI Max+ 395 processor at $N=4096$, fused peak-only filtering is $\ZenFleetPeakVsFFTW\times$ faster than a standalone AMD AOCL-FFTW inverse transform on a single CPU core, and $\GpuRoofNfourkPeakVsRoc\times$ faster than rocFFT on its integrated Radeon 8060S GPU; on the Apple M2 GPU it is $\MtwoNfourkPeakVsMlx\times$ faster than MLX FFT. Using a typical gravitational-wave search as a fiducial example (at a detection threshold of $\rho_\mathrm{th}=5.5$ and a $10^{-2}$ false-dismissal target), hierarchical screening yields a $\ZenFleetHierVsFFTW\times$ speedup over AOCL-FFTW on the CPU, $\GpuRoofNfourkHierVsRoc\times$ over rocFFT on the Radeon 8060S GPU, and $\MtwoNfourkHierVsMlx\times$ over MLX FFT on the Apple M2 GPU.
\end{abstract}

\begin{keyword}
Matched filtering \sep Operator fusion \sep Fast Fourier Transform \sep SIMD vectorization \sep Hierarchical search \sep Gravitational-wave astronomy
\end{keyword}

\end{frontmatter}

\section{Introduction}
\label{sec:intro}

\begin{figure*}[t]
  \centering
  \includegraphics[width=\textwidth]{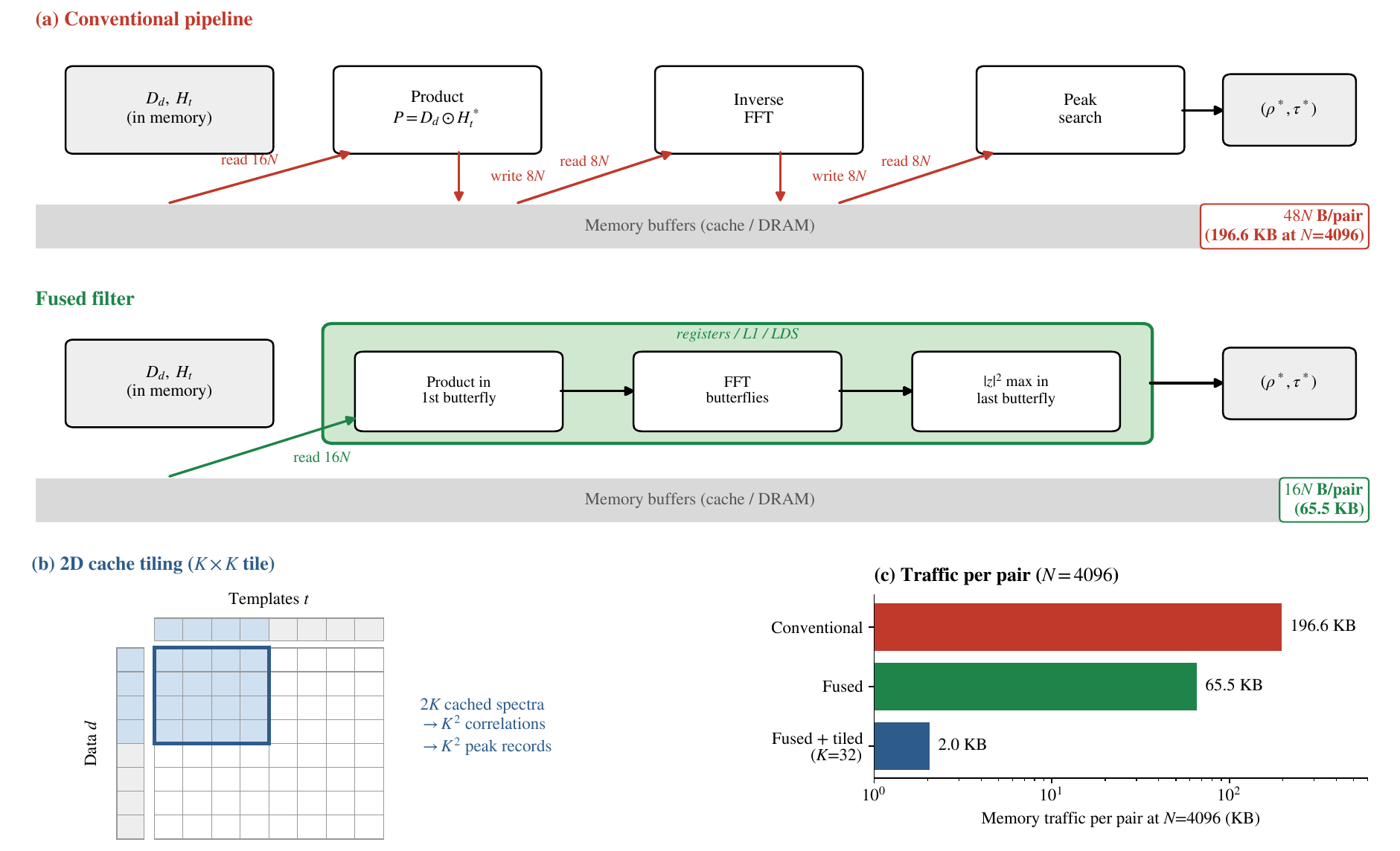}
  \caption{Memory traffic in unfused, fused, and tiled matched filtering. (a)~A conventional pipeline writes the product $P = D_d\odot H_t^*$ and the full correlation series $z[\tau]$ to memory buffers (cache or DRAM) and reads both back, transferring $48N$ bytes per pair. The fused filter applies the conjugate product in the first butterfly and the $|z|^2$ maximum in the last, keeping intermediate states in registers, L1 cache, or GPU local memory so only the $16N$ bytes of input spectra are read. (b)~In a $K\times K$ cache tile, $2K$ cached input spectra serve $K^2$ correlations. (c)~Memory traffic per pair at $N=4096$ (Eq.~\eqref{eq:amortized_traffic}): 196.6~KB conventional, 65.5~KB fused, and 2.0~KB fused and tiled with $K=32$.}
  \label{fig:dataflow}
\end{figure*}

Matched filtering is the optimal linear filter for maximizing the signal-to-noise ratio (SNR) of a known waveform in stationary Gaussian noise~\cite{allen2012findchirp,gibbons2007detection}. In a detection search algorithm, the parameters of the source are unknown, so searches correlate data against banks of discrete templates. This batched cross-correlation dominates the computational cost of scientific analyses, including compact-binary gravitational-wave searches~\cite{allen2012findchirp,usman2016pycbc,messick2017gstlal,nitz2018pycbclive,nitz2020ogc2}, repeating-event detection in seismology~\cite{gibbons2007detection}, and radio pulsar searches~\cite{ransom2002fourier,dimoudi2018gpu,adamek2020jerk}.

Direct time-domain correlation over $N$ samples naively requires $O(N^2)$ operations per data--template pair, whereas frequency-domain correlation via the Fast Fourier Transform (FFT) requires $O(N\log N)$ operations~\cite{cooley1965algorithm,frigo2005design}. In the frequency domain, for a batch of $n_d$ data segments and $n_t$ templates, each of the $n_d \times n_t$ pair evaluations reduces to an element-wise complex product, an inverse FFT (IFFT), and a scan for peaks above a detection threshold $\rho_\mathrm{th}$.

Although some search regimes naively appear to require long transforms ($N \sim 10^6$--$10^7$), whose working sets exceed on-chip memory and remain bandwidth-limited, long-duration correlations can often be recast into large batches of short transforms ($N \sim 10^3$--$10^5$). Fourier-domain pulsar acceleration sweeps~\cite{ransom2002fourier,dimoudi2018gpu,adamek2020overlap} and multi-channel seismic correlation detectors~\cite{gibbons2007detection} naturally operate on short-window transforms in this regime. Even in gravitational-wave searches, where early pipelines evaluated monolithic transforms of $N = 2^{18}$--$2^{22}$ samples~\cite{allen2012findchirp,usman2016pycbc}, modern formulations decompose long signals into shorter transforms via multi-band and multi-rate decompositions~\cite{cannon2012lloid,adams2016mbta}, adaptive multiband waveform evaluations~\cite{vinciguerra2017multibanding,morisaki2021multibanding}, and hierarchical or ratio-filter searches~\cite{dhurkunde2022hierarchical,nitz2026ratio}. At these transform lengths, individual transforms fit within on-chip cache or local memory, shifting the performance bottleneck to two sources of overhead: reading and writing intermediate arrays between separate pipeline kernels, and evaluating the full frequency band on pairs that contain only noise.

\begin{figure*}[t]
  \centering
  \includegraphics[width=\textwidth]{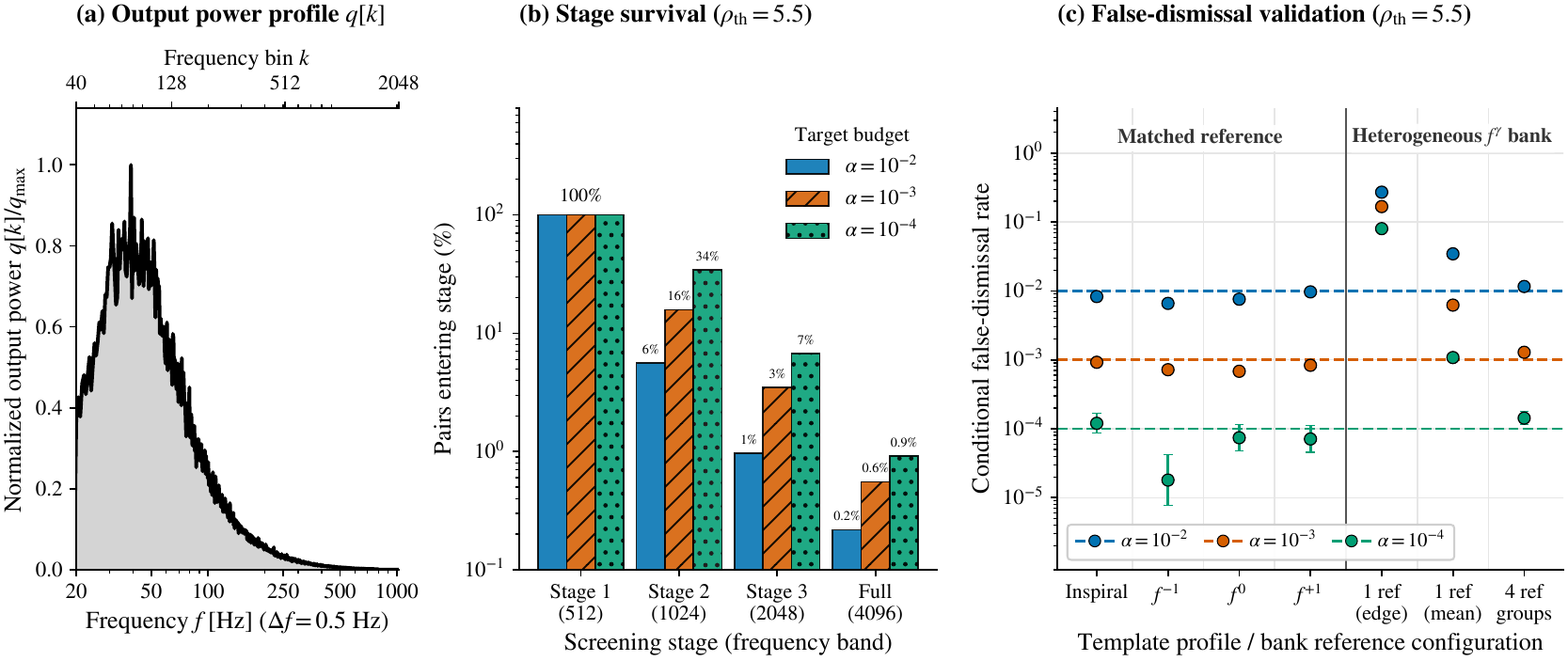}
  \caption{Statistically controlled hierarchical screening and empirical false-dismissal rate (FDR) validation at $N=4096$. (a)~Normalized matched-filter output power profile $q[k] = |H[k]|^2/S_n[k]$ vs.\ frequency $f$ ($\Delta f = 0.5~\mathrm{Hz}$, $f_\mathrm{low}=20~\mathrm{Hz}$ at bin $k=40$; top axis shows FFT bin $k$) for a PyCBC binary-neutron-star inspiral template whitened by an Advanced LIGO (aLIGO) noise curve. (b)~Percentage of data--template pairs entering each stage of the screening chain $(512, 1024, 2048)$ for $16\times512$ noise pairs at $\rho_\mathrm{th}=5.5$ across target budgets $\alpha \in \{10^{-2}, 10^{-3}, 10^{-4}\}$ (averaged over $5\times 10^5$ pairs); only $0.22\%$, $0.55\%$, and $0.91\%$ of pairs reach the full $N=4096$ filter. (c)~Measured conditional FDR with Wilson 95\% confidence intervals against targets $\alpha \in \{10^{-2}, 10^{-3}, 10^{-4}\}$ at $\rho_\mathrm{th}=5.5$: left of the divider shows synthetic templates matched to their own reference envelopes across inspiral, $f^{-1}$, $f^0$, and $f^{+1}$ spectra; right of the divider shows a heterogeneous power-law stress bank ($f^\gamma$, $\gamma \in [-2.33, -1.33]$) comparing a single edge-template reference, a single bank-mean reference, and four reference groups partitioned by spectral slope.}
  \label{fig:chain}
\end{figure*}

Several related domains have merged transform stages to reduce memory traffic. In code generation, SPIRAL~\cite{franchetti2005loop} fused stride permutations and scaling into the FFT computation. In machine learning, FFT-based convolution libraries introduced register and shared-memory GPU transforms~\cite{mathieu2014fast,vasilache2015fbfft}, and \texttt{FlashFFTConv}~\cite{fu2024flashfftconv} fused element-wise spectral products with the FFT and IFFT in on-chip memory, building on IO-aware tiling~\cite{dao2022flashattention}. In radio astronomy, \texttt{AstroAccelerate}~\cite{dimoudi2018gpu,adamek2020jerk,adamek2020overlap} fused the conjugate product and inverse FFT in GPU shared memory for Fourier-domain pulsar searches, but still wrote the full correlation power plane back to global memory for a separate candidate-selection pass, leaving peak extraction as a primary bottleneck~\cite{adamek2020jerk}. When a search requires only peaks above threshold, we can fuse both the conjugate product into the first stage of the inverse transform and the binned peak search into the final stage, avoiding the output time-series write entirely while tiling the $(n_d \times n_t)$ outer product in cache.

Even with operator fusion, evaluating the full transform on every pair wastes work when $>99\%$ of data--template pairs contain only background noise. Pruning candidates through coarse-to-fine cascades is widely used in sequential detection~\cite{sochman2005waldboost,rakthanmanon2012trillions}, sub-Nyquist radar~\cite{mishra2018subnyquist}, radio dispersion and pulsar searches~\cite{zackay2017fdmt,white2025pulscan}, and gravitational-wave searches~\cite{mohanty1996hierarchical,mohanty1998hierarchical,tanaka2000use,sengupta2002hierarchical,gadre2019hierarchical,soni2022hierarchical,dhurkunde2022hierarchical,nitz2026ratio}, alongside multi-rate and low-rank template decompositions~\cite{cannon2010svd,cannon2012lloid,messick2017gstlal,hooper2012spiir,adams2016mbta}.

Frequency-band screening is particularly effective when the expected SNR power $q[k]$ is concentrated at low frequencies while higher frequencies dominate the transform length $N$, as in gravitational-wave inspirals~\cite{tanaka2000use,sengupta2002hierarchical}. Existing hierarchical pipelines, however, treat coarse filtering as an external workflow stage with ad-hoc or heuristic thresholds tuned empirically via Monte Carlo signal injections~\cite{gadre2019hierarchical,soni2022hierarchical,dhurkunde2022hierarchical}, or interpolate and sum sub-band time series for every pair~\cite{adams2016mbta}. For the second source of overhead, we build multi-stage frequency-band screening directly into the matched-filter primitive itself and automatically determine optimal stage bandwidths and thresholds from the expected spectral power profile and matched-filter noise autocovariance $A \propto \mathcal{F}^{-1}\{q\}$ to satisfy a statistically defined target false-dismissal rate $\alpha$.

In this work we address both sources of overhead. First, we fuse the conjugate product into the first stage of the inverse transform and the peak search into the last, so that neither the product nor the correlation time series is written to memory, and we tile each batch so that a spectrum loaded into cache serves many data--template pairs. Second, we screen pairs with a short sequence of low-frequency transforms before evaluating the full band. The threshold of each stage is set from the expected signal power spectrum and the matched-filter noise autocovariance, so that the fraction of signals lost meets a chosen false-dismissal target. We implement both methods in the open-source \texttt{matchedfilter} library and benchmark them on eight CPU architectures and two integrated GPUs.

Section~\ref{sec:background} reviews Fourier-domain correlation, the FFT, and the memory traffic of a conventional pipeline. Section~\ref{sec:fusion} describes our operator-fused correlation kernels and 2D batch tiling. Section~\ref{sec:prefilter} derives the covariance-based screening model and tests its false-dismissal accuracy. Section~\ref{sec:benchmarks} presents CPU and GPU performance benchmarks, and Section~\ref{sec:conclusion} summarizes our conclusions.

\section{Background}
\label{sec:background}
\label{sec:formulation}

Matched-filtering applications involve computing correlations between data and template signals. Consider a set of $n_d$ frequency-domain data segments\footnote{We ignore the cost of transforming the data/template into the frequency domain if either were originally sampled in time, as this can be done once up front.} $D_d$ ($d \in [0, n_d-1]$), each of length $N$. These are correlated with a set of
$n_t$ frequency-domain templates $H_t$ ($t \in [0, n_t-1]$), which are also of length $N$. The complex correlation time series for a $(d, t)$ data-segment, template pair at lag $\tau$ is\footnote{Here, we assume that the data consists of a signal plus white noise. If the noise is colored, the data and template can be whitened without loss of generality by letting $D_d[k] \leftarrow D_d[k]/\sqrt{S_n[k]}$, $H_t[k] \leftarrow H_t[k]/\sqrt{S_n[k]}$, where $S_n[k]$ is the power spectral density of the noise at index $k$, with $H_t$ normalized such that $\mathbb{E}[|z_{dt}[\tau]|^2] = 1$ in pure noise~\cite{allen2012findchirp}.}
\begin{equation}
z_{dt}[\tau] = \frac{1}{N}\!\sum_{k=0}^{N-1} D_d[k] H_t^*[k] e^{2\pi i k \tau / N} = \mathcal{F}^{-1}\!\left\{ D_d \odot H_t^* \right\}\![\tau].
\label{eq:mf}
\end{equation}
The signal-to-noise ratio (SNR) time series of the data/template pair $\rho_{dt}[\tau]$ is the magnitude of $z_{dt}[\tau]$. In stationary Gaussian noise, $\rho_{dt}$ is the optimal detection statistic for a signal of known shape and unknown phase, so candidate signals appear as peaks in $\rho_{dt}[\tau]$~\cite{allen2012findchirp}. Many searches therefore keep only the peak magnitude and lag within discrete time intervals of width $N_w$,
\begin{equation}
\rho_{dt}[b] = \max_{\tau \in [b N_w,\, (b+1) N_w - 1]} |z_{dt}[\tau]|, \quad b \in \{0, \dots, N/N_w - 1\}.
\label{eq:binned_peak}
\end{equation}
This reduces to a single peak record $(\rho_{dt}^*, \tau_{dt}^*)$ per $(d,t)$ pair when $N_w = N$. Below we drop the $dt$ subscript where the pair is clear from context.

An $N$-point FFT with $N = 2^p$ evaluates Eq.~\eqref{eq:mf} in $p = \log_2 N$ stages. Each stage combines pairs of complex values $(a, b)$ into $(a + wb,\, a - wb)$, where the twiddle factor $w$ is a root of unity; this two-input, two-output step is called a butterfly. A stage performs $N/2$ butterflies of one complex multiplication and two complex additions each (10 real operations), giving the usual count of $5N\log_2 N$ operations~\cite{cooley1965algorithm,frigo2005design}. In a standalone library call, the first stage reads its inputs from memory and the last stage writes its outputs to memory. The first stage is also the first point at which each input is needed, and the last stage the first point at which each output is final. We can therefore apply the conjugate product as the first stage reads its inputs and compare against the running peak as the last stage produces its outputs, without storing either array. The same holds for the four-step decomposition we use for longer transforms (Section~\ref{sec:fusion}).

Conventional correlation pipelines across these domains~\cite{allen2012findchirp,usman2016pycbc,nitz2018pycbclive,gibbons2007detection,ransom2002fourier} evaluate Eqs.~\eqref{eq:mf}--\eqref{eq:binned_peak} by calling a standalone FFT library~\cite{frigo2005design,tolmachev2023vkfft} between separate multiplication and peak-finding kernels:
\begin{equation*}
(D_d, H_t) \;\to\; P = D_d \odot H_t^* \;\to\; z = \operatorname{IFFT}(P) \;\to\; \max_\tau |z[\tau]|.
\end{equation*}
In single precision (8 bytes per complex sample), this sequence moves six length-$N$ complex arrays through the memory hierarchy per
pair: reading $D_d$ and $H_t$ ($16N$ bytes), writing and re-reading the product $P$ ($16N$ bytes) and time series $z$ ($16N$ bytes), for a total of $48N$ bytes per pair. Since a radix-2 correlation executes roughly $5N\log_2 N + 6N$ floating-point operations ($5N\log_2 N$ for the transform and $6N$ for the conjugate product~\cite{cooley1965algorithm,frigo2005design}) while moving $48N$ bytes, its arithmetic intensity is only
\begin{equation}
I_\mathrm{unfused} = \frac{5 \log_2 N + 6}{48} \approx 1.38 \; \frac{\text{FLOP}}{\text{byte}} \quad (N=4096),
\label{eq:unfused_ai}
\end{equation}
far below the compute--memory balance point ($20$--$150\text{ FLOP/byte}$) of modern processors~\cite{williams2009roofline,dao2022flashattention}. Even when an individual array fits in L2 or L3 cache, writing and re-reading $P$ and $z$ consumes cache capacity and bandwidth on every pair and spills to main memory across batched evaluations.

\section{Operator-Fused Correlation and Batch Tiling}
\label{sec:fusion}

\begin{figure*}[t]
  \centering
  \includegraphics[width=\textwidth]{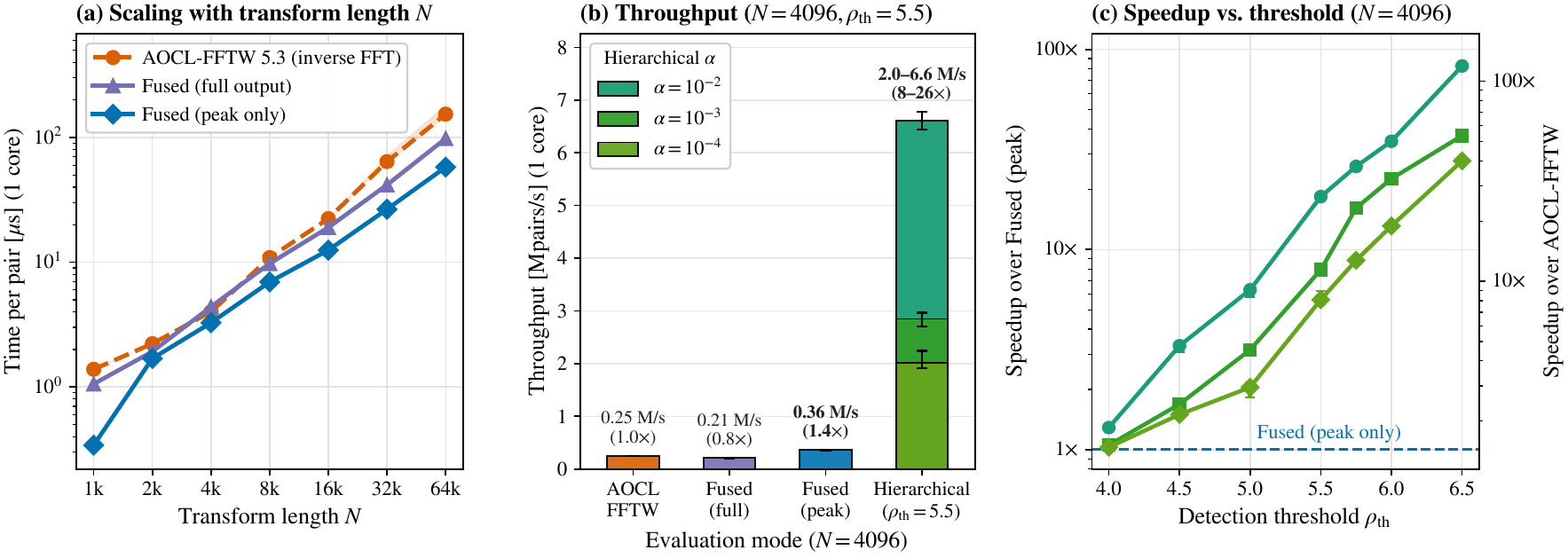}
  \caption{Single-core CPU performance on the AMD Ryzen AI Max+ 395 (Zen~5, $16 \times 512 = 8{,}192$ pairs per batch). (a)~Time per pair across transform lengths $N \in [1024, 65536]$ for AMD AOCL-FFTW 5.3 (\texttt{FFTW\_PATIENT}, batched inverse transform only) and the fused full-output and peak-only filters; shaded bands span the 10th--90th percentile of timing blocks. (b)~Throughput in millions of pairs per second at $N=4096$ and $\rho_\mathrm{th}=5.5$ in Gaussian noise with the PyCBC binary-neutron-star reference profile, showing AOCL-FFTW, fused full-output, fused peak-only, and hierarchical screening with false-dismissal budgets $\alpha \in \{10^{-4}, 10^{-3}, 10^{-2}\}$ overlaid (speedup over AOCL-FFTW in parentheses). (c)~Speedup of hierarchical screening over fused peak-only filtering (left axis) and over AOCL-FFTW (right axis) as a function of the detection threshold $\rho_\mathrm{th} \in [4.0, 6.5]$ at $N=4096$.}
  \label{fig:cpu_latency}
\end{figure*}

Figure~\ref{fig:dataflow} illustrates how operator fusion and 2D batch tiling reduce memory traffic. We fuse the conjugate multiplication $D_d \odot H_t^*$ into the first butterfly stage of the inverse transform and the binned peak search $\max_\tau |z[\tau]|^2$ into the final butterfly stage. All intermediate butterfly states remain in vector registers, L1 cache, or GPU local data share ($8N$ bytes per active transform, or $32~\mathrm{KB}$ at $N=4096$). Removing the intermediate product and output time-series buffers eliminates $32N$ bytes/pair of memory writes and reads, leaving only the $16N$-byte input spectral read and a 16-byte peak record per pair.

This fusion of the input product and output peak reduction is independent of the internal FFT decomposition. In our implementation, short transforms that fit entirely within L1 cache on the CPU ($N \le 1024$) evaluate independent data--template pairs in parallel across SIMD lanes, whereas larger transforms ($N \ge 2048$ on CPU and all GPU transforms) use a four-step $N = N_1 \times N_2$ decomposition~\cite{cooley1965algorithm,bailey1990ffts} with inputs pre-permuted once upon ingest into unit-stride split real and imaginary arrays ($<2.5\%$ of batch runtime). In all cases, the final butterfly stage tracks the maximum squared magnitude $|z[\tau]|^2$ across each time bin in registers and takes a single square root per bin to emit $(\rho^*, \tau^*)$; when full correlation series are required downstream, a companion full-output kernel streams $z[\tau]$ at the end of the transform while still avoiding the intermediate product buffer $P$.

When correlating $n_d$ data segments against a large template bank ($n_t \gg 10^3$), looping over all $n_t$ templates for each data segment streams the entire bank through cache $n_d$ times, evicting templates whenever $n_t \times 8N$ exceeds cache capacity $C_\mathrm{cache}$. We instead partition the $n_d \times n_t$ outer product into 2D $(n_{d,\mathrm{tile}} \times n_{t,\mathrm{tile}})$ tiles chosen so that $(n_{d,\mathrm{tile}} + n_{t,\mathrm{tile}}) \times 8N \le C_\mathrm{cache}$ (Figure~\ref{fig:dataflow}(b)). For a $K \times K$ tile, the $K$ data spectra and $K$ template spectra ($2K$ spectra, $16KN$ bytes)
are loaded once and serve all $K^2$ pairs in the tile, which write $K^2$ 16-byte peak records:
\begin{equation}
M_\mathrm{tiled}(K) = 16KN + 16K^2~\mathrm{B}, \quad
M_\mathrm{pair} = \frac{16N}{K} + 16~\mathrm{B/pair}.
\label{eq:amortized_traffic}
\end{equation}%
Relative to the $48N$ bytes/pair of an unfused pipeline, fused 2D tiling reduces memory traffic outside the cache tile by
\begin{equation}
\mathcal{R} = \frac{48N K^2}{16NK + 16K^2} \approx 3K\times
\label{eq:reuse_factor}
\end{equation}
($\sim 100\times$ at $K=32, N=4096$, or $2.0~\mathrm{KB}$/pair, Figure~\ref{fig:dataflow}(c)), lifting arithmetic intensity above $100\text{ FLOP/byte}$.

\section{Statistically Controlled Hierarchical Screening}
\label{sec:prefilter}

\subsection{Signal Power Envelope and Noise Covariance Model}
\label{sec:cov_model}
To avoid evaluating the full-bandwidth transform on noise-only pairs, a coarse screening stage can evaluate the correlation on the first $m < N$ frequency bins and run the full filter only when the coarse peak exceeds a threshold $\rho_c$~\cite{tanaka2000use,sengupta2002hierarchical}. We describe each template by its expected matched-filter output power at frequency bin $k$, $q[k] = |H[k]|^2/S_n[k]$ (equivalently $|H[k]|^2$ for a whitened template). Setting $\rho_c$ requires accounting for three effects: (i) truncating the band to $m$ bins retains only a fraction of the signal power,
\begin{equation}
\mu = \frac{\sum_{k=0}^{m-1} q[k]}{\sum_{k=0}^{N-1} q[k]};
\end{equation}%
(ii) evaluating the coarse correlation on a decimated lag grid of spacing $N/m$ introduces scalloping loss when the true peak falls between coarse samples~\cite{ransom2002fourier}; and (iii) band truncation changes the correlation length of the filtered noise.

For a template with within-band normalized output-power weights $\hat{q}[k] = q[k] / \sum_{j=0}^{m-1} q[j]$ and a signal with full-band SNR $\rho$ arriving at lag $L$ (measured in units of the coarse sample spacing $N/m$, and in general not an integer), the expected unit-variance coarse response $z_c[\tau]$ at coarse lag $\tau$ is
\begin{equation}
\begin{aligned}
\mathbb{E}[z_c[\tau]] &= \rho \sqrt{\mu} \, A(\tau - L), \\
A(x) &= \sum_{k=0}^{m-1} \hat{q}[k] \, e^{2\pi i k x / m},
\end{aligned}
\label{eq:signal_response}
\end{equation}
with $A(0) = 1$. We write $A$ with parentheses because it is evaluated at non-integer offsets $\tau - L$. In stationary Gaussian noise, the temporal autocovariance of the unit-variance filtered noise in that band is given by the same inverse transform~\cite{allen2012findchirp}:
\begin{equation}
\mathbb{E}\!\left[ n[\tau_1] n^*[\tau_2] \right] = A(\tau_1 - \tau_2) = m\,\mathcal{F}^{-1}\{\hat{q}\}[\tau_1 - \tau_2],
\label{eq:covariance}
\end{equation}
for coarse lags $\tau_1, \tau_2$, where the factor $m$ cancels the $1/m$ normalization of the length-$m$ inverse transform (Eq.~\eqref{eq:mf}). Note that $A$ depends only on the spectral power profile $q[k]$ and not on the phase evolution $\arg H[k]$ that distinguishes individual templates. Once computed for a reference power profile, the coarse signal response, noise covariance, and resulting screening thresholds can be reused across all templates with a similar spectral envelope.

\subsection{Multi-Stage Screening Chains and Cost Selection}
\label{sec:chain}

Rather than relying on a single coarse band, we chain up to three coarse tiers at increasing power-of-two bands $b_1 < b_2 < b_3 < N$. Tier $i$ correlates the first $b_i$ frequency bins on a lag grid of spacing $N/b_i$ and passes only pairs exceeding $\rho_{c,i}$ to tier $i+1$; survivors of the final tier run the full length-$N$ fused filter. Each tier is the single-band model of Section~\ref{sec:cov_model} with $m = b_i$; we write $\mu_i$ and $A_{b_i}$ for the corresponding $\mu$ and $A$.

For independent whitened frequency bins, the cross-covariance between unit-variance filtered noise in bands $b_i \le b_j$ at a time separation $x$, measured in units of the band-$b_i$ sample spacing $N/b_i$, is $\sqrt{\mu_i / \mu_j}\, A_{b_i}(x)$. Under this joint covariance, the conditional false-dismissal rate (FDR) is the probability that a signal reaching $\rho_\mathrm{th}$ in the full filter is rejected by any screening tier:
\begin{equation}
P_\mathrm{FD}(\boldsymbol{\rho}_c) = P\!\left( \bigcup_{i} \left\{\max_{\tau \in \Omega_{c,i}} |z_{c,i}[\tau]| < \rho_{c,i}\right\} \;\middle|\; \max_{\tau \in \Omega_f} |z_f[\tau]| \ge \rho_\mathrm{th} \right),
\label{eq:p_fd}
\end{equation}
where $z_{c,i}$ and $z_f$ are the tier-$i$ and full-band correlations and $\Omega_{c,i}$ and $\Omega_f$ are the sets of lags searched in the block on the tier-$i$ and full-rate grids. Since $A$ is localized around the true peak, we sample correlated Gaussian realizations via Cholesky factorization over a local window around the peak across sub-sample arrival offsets, combine each draw's local coarse maximum with the noise maximum over the remainder of the searched block, and set the tier thresholds $\boldsymbol{\rho}_c$ by quantile inversion so that $P_\mathrm{FD}(\boldsymbol{\rho}_c) = \alpha$.

To select the fastest chain $(b_1, \dots, b_k)$ and allocate the false-dismissal budget $\alpha$ across its tiers on a given processor, the library benchmarks the per-pair tier costs $c_i$ once at initialization and minimizes the expected cost against the modeled cumulative noise survival fractions $\pi_i$:
\begin{equation}
C = c_1 + \sum_{i=2}^{k} c_i \, \pi_{i-1} + c_\mathrm{fine} \, \pi_k,
\label{eq:chain_cost}
\end{equation}
timing the top candidate chains on a trial batch to choose the winner. Cholesky threshold calibration and chain selection run once at plan creation in $\sim 0.1~\mathrm{s}$. As shown in Figure~\ref{fig:chain}(a)--(b), for a binary-neutron-star template whitened by an Advanced LIGO (aLIGO) noise curve at $N=4096$ and $\rho_\mathrm{th}=5.5$, the first 512 bins ($f \le 256~\mathrm{Hz}$) already capture $96.5\%$ of the full-band SNR ($\sqrt{\mu} = 0.965$, rising to $99.3\%$ at band 1024 and $100\%$ at band 2048, above the $800~\mathrm{Hz}$ template cutoff). The selected chain $(512, 1024, 2048)$ rejects $94.4\%$ ($\alpha=10^{-2}$), $84.2\%$ ($\alpha=10^{-3}$), and $65.7\%$ ($\alpha=10^{-4}$) of noise pairs in Stage~1, leaving only $0.22\%$, $0.55\%$, and $0.91\%$ of pairs to be evaluated by the full filter. The rejection rates hold for noise-dominated data which is approximately Gaussian; rejection rates would be more permissive where these assumptions do not hold.

\subsection{Empirical False-Dismissal Validation and Reference Groups}
\label{sec:fdr_validation}
Figure~\ref{fig:chain}(c) tests the empirical false-dismissal rate of the screening model using simulated data with Gaussian noise at $N=4096$ and $\rho_\mathrm{th}=5.5$. A false dismissal is recorded whenever an injection reaches $\rho \ge \rho_\mathrm{th}$ in the full-bandwidth filter, but is rejected by the screening chain.

When templates match their reference envelope across four spectral profiles (a binary-neutron-star inspiral profile alongside $f^{-1}$, $f^0$, and $f^{+1}$ power laws; left side of Figure~\ref{fig:chain}(c), $>2.5\times 10^5$ detected injections per profile), the measured conditional FDR closely tracks the target $\alpha \in \{10^{-2}, 10^{-3}, 10^{-4}\}$ across all three decades. In a standard binary-neutron-star search, templates share a nearly universal $f^{-7/3}$ amplitude profile prior to merger, so a single reference profile suffices across the bank.

When a single template bank spans a wide range of spectral slopes, a single reference envelope is no longer sufficient. The right side of Figure~\ref{fig:chain}(c) evaluates a template bank where the templates span a range of spectral slopes $f^\gamma$ with $\gamma \in [-2.33, -1.33]$. Using the steepest template ($\gamma = -2.33$), or the bank mean as a single reference for the entire bank, overestimates the low-frequency power fraction of flatter templates, causing elevated false dismissals ($\sim 27\%$ bank-wide for the edge reference and $\sim 3.5\%$ for the mean reference at $\alpha=10^{-2}$). Partitioning the bank into four reference groups ordered by spectral slope restores target compliance across the bank ($1.2 \times 10^{-2}$, $1.3 \times 10^{-3}$, and $1.4 \times 10^{-4}$ for $\alpha = 10^{-2}, 10^{-3}$, and $10^{-4}$, respectively). Therefore, template banks whose in-band power fractions vary significantly across the parameter space should be partitioned into reference groups with similar spectral envelopes, or may require upfront evaluation of the thresholds on a template-by-template basis.

\section{Performance and Numerical Evaluation}
\label{sec:benchmarks}

\begin{table}[t]
\centering
\caption{Single-precision fused peak-only filtering against a double-precision (\texttt{float64}) reference: relative peak-magnitude error $\delta_\rho$ per transform length over 640 trials per $N$ (128 each at injected SNR 0, 5, 8, 20, 100; SNR 0 is noise only). All trials show complete agreement on the peak sample index.}
\label{tab:numerical_accuracy}
\footnotesize
\setlength{\tabcolsep}{5pt}
\begin{tabular}{rccc}
\toprule
$N$ & Median $\delta_\rho$ & 99th pct.\ $\delta_\rho$ & Max.\ $\delta_\rho$ \\
\midrule
1,024 & $4.26 \times 10^{-8}$ & $1.47 \times 10^{-7}$ & $1.76 \times 10^{-7}$ \\
2,048 & $4.73 \times 10^{-8}$ & $1.58 \times 10^{-7}$ & $1.82 \times 10^{-7}$ \\
4,096 & $4.70 \times 10^{-8}$ & $1.46 \times 10^{-7}$ & $1.78 \times 10^{-7}$ \\
8,192 & $4.61 \times 10^{-8}$ & $1.54 \times 10^{-7}$ & $1.74 \times 10^{-7}$ \\
16,384 & $4.69 \times 10^{-8}$ & $1.56 \times 10^{-7}$ & $1.92 \times 10^{-7}$ \\
32,768 & $5.36 \times 10^{-8}$ & $1.70 \times 10^{-7}$ & $2.26 \times 10^{-7}$ \\
65,536 & $7.10 \times 10^{-8}$ & $1.80 \times 10^{-7}$ & $2.30 \times 10^{-7}$ \\
\bottomrule
\end{tabular}
\end{table}

\begin{table*}[t]
\centering
\small
\setlength{\tabcolsep}{5pt}
\caption{CPU time per batch (ms; one pinned thread; $16 \times 512$ pairs, $N=4096$; median of timing blocks). The FFTW column times only the batched inverse transform (using AMD AOCL-FFTW on the Ryzen AI Max+ 395 and FFTW on the remaining hosts). Hierarchical columns use $\rho_\mathrm{th}=5.5$. Last three columns: speedup of peak-only and hierarchical ($\alpha=10^{-2}$) filtering over FFTW, and of hierarchical ($\alpha=10^{-2}$) over peak-only. $^\dagger$Shared or contended host; the Xeon 8260 is a virtual machine.}
\label{tab:fleet_cpu}
\begin{tabular}{lrrrrrrrrr}
\toprule
CPU & FFTW & Full & Peak & \multicolumn{3}{c}{Hierarchical, $\alpha =$} & \multicolumn{2}{c}{vs.\ FFTW} & Hier. \\
\cmidrule(lr){5-7}\cmidrule(lr){8-9}
 & & & & $10^{-2}$ & $10^{-3}$ & $10^{-4}$ & Peak & Hier. & vs.\ Peak \\
\midrule
Ryzen AI Max+ 395 (Zen 5) & 32.84 & 39.03 & 22.78 & 1.24 & 2.88 & 4.06 & 1.4 & 26 & 18 \\
EPYC 9845 & 83.44 & 41.96 & 29.64 & 2.52 & 4.67 & 5.48 & 2.8 & 33 & 12 \\
Ryzen 9 5950X (Zen 3) & 64.76 & 52.43 & 34.82 & 2.20 & 4.42 & 5.67 & 1.9 & 29 & 16 \\
Core i5-13500H & 45.71 & 56.49 & 36.09 & 2.30 & 4.30 & 6.71 & 1.3 & 20 & 16 \\
Ryzen 5 5500U (Zen 2) & 88.53 & 73.49 & 48.80 & 3.66 & 7.19 & 11.29 & 1.8 & 24 & 13 \\
Apple M2 & 70.61 & 69.95 & 58.07 & 3.24 & 5.54 & 7.70 & 1.2 & 22 & 18 \\
Xeon Platinum 8260 (VM)$^\dagger$ & 218.47 & 127.20 & 78.89 & 6.68 & 12.36 & 14.87 & 2.8 & 33 & 12 \\
Xeon E5-2698 v3$^\dagger$ & 320.55 & 257.70 & 139.74 & 11.35 & 20.08 & 28.59 & 2.3 & 28 & 12 \\
\bottomrule
\end{tabular}
\end{table*}

We evaluate single-core scaling, numerical accuracy, and hierarchical screening on an AMD Ryzen AI Max+ 395 processor (Zen~5, AVX-512), a cross-platform fleet of eight CPU architectures, and two integrated GPUs (an AMD Radeon 8060S via Vulkan and an Apple M2 GPU via Metal). All benchmarks measure steady-state execution on batches of $n_d=16$ data segments $\times$ $n_t=512$ templates ($8{,}192$ pairs), excluding one-time plan creation.

\subsection{CPU Performance and Numerical Accuracy}
\label{sec:cpu_bench}
\label{sec:numerical}
\label{sec:fleet}

Figure~\ref{fig:cpu_latency}(a) compares the fused full-output and peak-only filters on one pinned Zen~5 core against AMD AOCL-FFTW 5.3 (\texttt{FFTW\_PATIENT}). The AOCL-FFTW and stock FFTW baselines time only the standalone batched inverse transform without the $n_d \times n_t$ conjugate multiplications ($D_d \odot H_t^*$) or peak search. The fused full-output kernel performs both the conjugate product and inverse transform in a single pass while still writing the full correlation time series ($256~\mathrm{MB}$ per batch at $N=4096$) to main memory. The fused peak-only kernel eliminates this output write by identifying and collecting peaks in registers. Across $N \in [1024, 65536]$, fused peak-only filtering is $\ZenScalePeakVsAoclmin$--$\ZenScalePeakVsAoclmax\times$ faster than standalone AOCL-FFTW. At $N=1024$, where an entire transform fits within L1 cache, the cross-pair SIMD kernel reaches $\ZenScalePeakA~\mu\mathrm{s}$ per pair ($\ZenScalePeakVsAoclmax\times$ faster than AOCL-FFTW); for $N \ge 2048$, the four-step kernel vectorizes within each transform, maintaining a $\ZenScalePeakVsAoclmin$--$2.7\times$ speedup over AOCL-FFTW, where the separation between the full-output and peak-only curves directly measures the saved memory write of the length-$N$ correlation series.

Figure~\ref{fig:cpu_latency}(b)--(c) shows the end-to-end throughput and threshold scaling when adding hierarchical screening at $N=4096$ for the binary-neutron-star inspiral profile in Gaussian noise. At $\rho_\mathrm{th}=5.5$ (Figure~\ref{fig:cpu_latency}(b)), single-core throughput increases from $0.25~\mathrm{Mpairs/s}$ ($\ZenFleetFFTW~\mathrm{ms}$ per batch) for a standalone AOCL-FFTW inverse transform to $0.36~\mathrm{Mpairs/s}$ ($\ZenFleetPeak~\mathrm{ms}$, $\ZenFleetPeakVsFFTW\times$) for fused peak-only filtering, and to $2.0$--$6.6~\mathrm{Mpairs/s}$ ($\ZenFleetHier$--$4.06~\mathrm{ms}$) for hierarchical screening across $\alpha \in [10^{-4}, 10^{-2}]$ ($18\times$ over fused peak-only filtering and $\ZenFleetHierVsFFTW\times$ over AOCL-FFTW at $\alpha=10^{-2}$). As shown in Figure~\ref{fig:cpu_latency}(c), hierarchical screening reaches break-even with fused peak-only filtering near $\rho_\mathrm{th}=\FleetThrLo$ ($\FleetThrLoMin$--$\FleetThrLoMax\times$), where background noise peaks frequently exceed the coarse threshold, and rises smoothly with detection threshold to $6$--$18\times$ at $\rho_\mathrm{th}=5.5$ and $\FleetThrHiMin$--$\FleetThrHiMax\times$ at $\rho_\mathrm{th}=\FleetThrHi$. In the rare-signal regime, throughput is dominated by the noise. Hierarchical screening would become less efficient only if the signal rate were high enough to significantly increase the overall rate of excursions above the detection threshold.

Table~\ref{tab:numerical_accuracy} compares the single-precision (\texttt{float32}) fused peak-only filter against a double-precision (\texttt{float64}) reference across $\AccTrials$ trials per transform length spanning injected SNRs from $0$ (noise only) to $100$. The median relative peak error $\delta_\rho = |\rho_\mathrm{fused} - \rho_\mathrm{ref}|/\rho_\mathrm{ref}$ grows slowly from $\AccMedianFirst$ at $N=\AccNfirst$ to $\AccMedianLast$ at $N=\AccNlast$, with a maximum error across all $4{,}480$ trials of $\AccMax$ ($\sim 4\times$ single-precision machine epsilon $\epsilon = 2^{-24}$) and complete agreement on the peak sample index in every trial.

Table~\ref{tab:fleet_cpu} extends the $N=4096$ benchmark across eight CPUs spanning x86 (AVX2 and AVX-512) and ARM64 (Apple M2 NEON). Across all tested processors, fused peak-only filtering outperforms a standalone inverse FFT by $\FleetPeakVsFFTWmin$--$\FleetPeakVsFFTWmax\times$, while hierarchical screening at $\rho_\mathrm{th}=5.5, \alpha=10^{-2}$ achieves an additional $\FleetHierVsPeakmin$--$\FleetHierVsPeakmax\times$ speedup over peak-only filtering ($\FleetHierVsFFTWmin$--$\FleetHierVsFFTWmax\times$ total speedup over FFTW). Tightening the false-dismissal budget by two orders of magnitude, from $\alpha=10^{-2}$ to $10^{-4}$, increases runtime by only $\sim 2$--$3\times$ across the machines tested.

\subsection{GPU Roofline and Memory-Wall Analysis}
\label{sec:gpu_bench}

\begin{figure*}[!t]
  \centering
  \includegraphics[width=\textwidth]{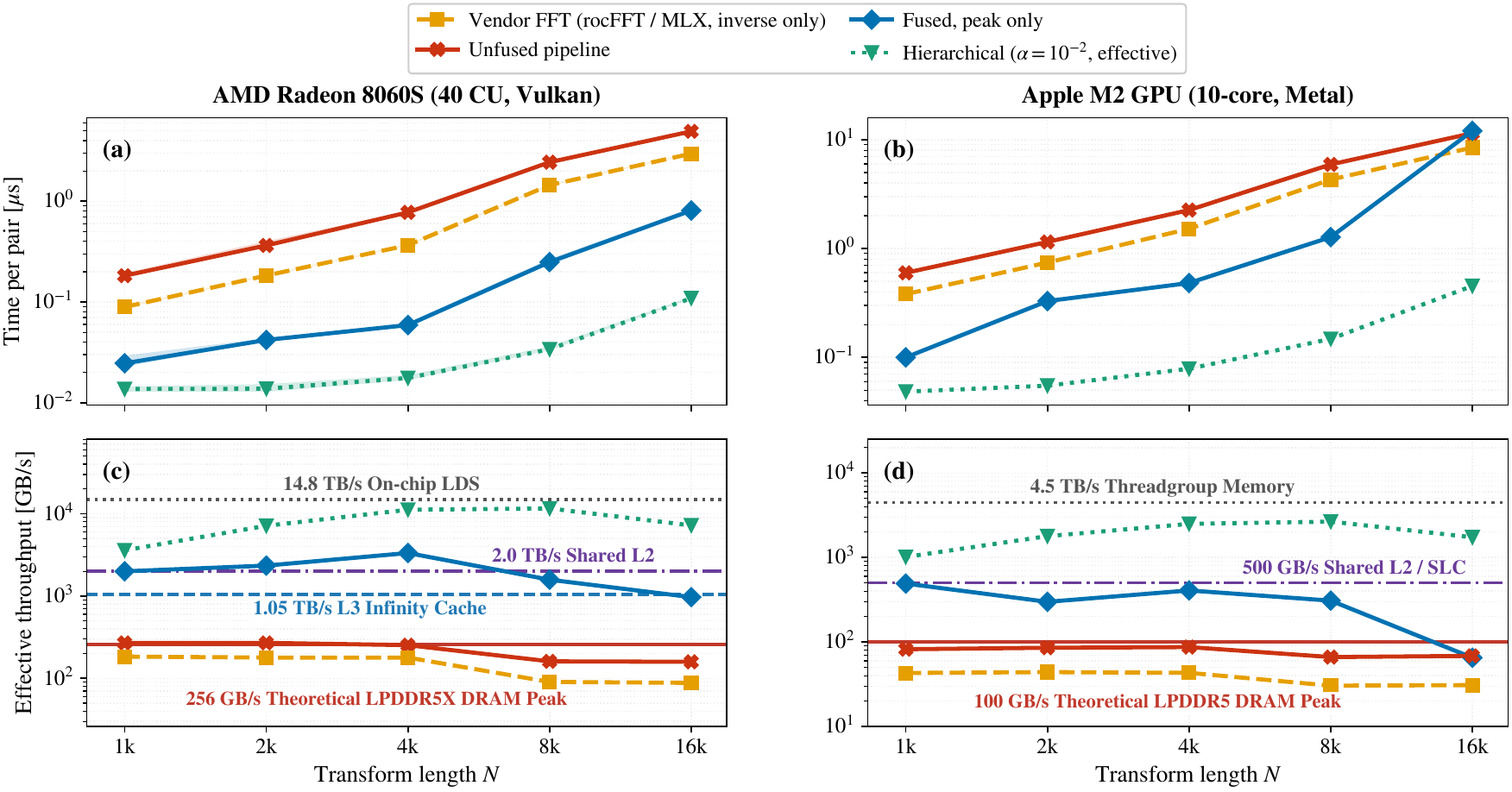}
  \caption{GPU performance and effective problem throughput on the integrated AMD Radeon 8060S (40~CUs, Vulkan; left) and Apple M2 GPU (10~cores, Metal; right) for $16 \times 512 = 8{,}192$ pairs per batch. (a,~b)~Time per pair, including device synchronization, for the vendor FFT baseline (rocFFT or MLX inverse transform only), the complete unfused pipeline (conjugate product $\to$ inverse FFT $\to$ single-pass peak search), and the fused peak-only and hierarchical screening filters ($\rho_\mathrm{th}=5.5, \alpha=10^{-2}$). (c,~d)~Effective problem throughput, defined as $48N$ conventional problem bytes divided by execution time ($16N$ for the standalone inverse FFT), compared against each device's memory hierarchy ceilings: LPDDR5X DRAM (256~GB/s), Infinity Cache (1.05~TB/s), shared L2 (2.0~TB/s), and on-chip Local Data Share (LDS, 14.8~TB/s) on the Radeon 8060S; and LPDDR5 DRAM (100~GB/s), shared L2/System Level Cache (SLC, 500~GB/s), and on-chip threadgroup memory (4.5~TB/s) on the Apple M2. The hierarchical curves in panels~(c) and~(d) are shown only as equivalent-rate performance references and do not reflect physical memory throughput, as screening skips operations and memory reads on rejected pairs. The throughput is bound by DRAM bandwidth without operation fusion and generally bound by L2 bandwidth for the fused, peak-only version.}
  \label{fig:gpu_roofline}
\end{figure*}

Figure~\ref{fig:gpu_roofline} evaluates the same $16 \times 512$ workload on two integrated GPUs spanning distinct graphics APIs and memory hierarchies: the AMD Radeon 8060S (40~CUs, Vulkan, 256~GB/s LPDDR5X) and the Apple M2 GPU (10~cores, Metal, 100~GB/s LPDDR5). On the Radeon 8060S across $N \in [1024, 16384]$ (Figure~\ref{fig:gpu_roofline}(a)), fused peak-only filtering is $\GpuRoofPeakVsUnfusedmin$--$\GpuRoofPeakVsUnfusedmax\times$ faster than the complete unfused HIP + rocFFT pipeline and $\GpuRoofPeakVsRocmin$--$\GpuRoofPeakVsRocmax\times$ faster than a standalone rocFFT inverse transform, while hierarchical screening ($\rho_\mathrm{th}=5.5, \alpha=10^{-2}$) is $\GpuRoofHierVsRocmin$--$\GpuRoofHierVsRocmax\times$ faster than rocFFT. At $N=4096$, batch time on the 8060S drops from $\GpuRoofNfourkRoc~\mathrm{ms}$ (rocFFT inverse only) and $\GpuRoofNfourkFull~\mathrm{ms}$ (fused full-output) to $\GpuRoofNfourkPeak~\mathrm{ms}$ for fused peak-only filtering ($\GpuRoofNfourkPeakVsRoc\times$ over rocFFT) and $\GpuRoofNfourkHier~\mathrm{ms}$ for hierarchical screening at $\alpha = 10^{-2}$ ($\GpuRoofNfourkHierVsRoc\times$ over rocFFT). On the Apple M2 GPU (Figure~\ref{fig:gpu_roofline}(b)), fused peak-only filtering for $N \le 8192$ is $3.5$--$6.0\times$ faster than the complete unfused pipeline (MLX conjugate product $\to$ MLX inverse FFT $\to$ Metal peak search) and $2.3$--$3.8\times$ faster than a standalone MLX inverse transform, while hierarchical screening is $\MtwoHierVsMlxmin$--$\MtwoHierVsMlxmax\times$ faster than a standalone MLX inverse transform across $N \in [1024, 16384]$.

Plotting effective problem throughput ($48N$ bytes divided by execution time, Figure~\ref{fig:gpu_roofline}(c)--(d)) shows the role of the GPU memory hierarchy on both architectures. On the Radeon 8060S (panel~(c)), once the intermediate buffers spill out of the 32~MB Infinity Cache (L3), the unfused pipeline is bound by the LPDDR5X DRAM ceiling ($\GpuRoofUnfusedBWmax \to \GpuRoofUnfusedBWmin\text{ GB/s}$, matching the $\GpuRoofCopy\text{ GB/s}$ measured copy limit), whereas fused peak-only filtering keeps butterfly states inside shared L2 cache and on-chip Local Data Share (LDS), sustaining $\sim 1$--$3.3\text{ TB/s}$ of effective throughput. The Apple M2 GPU (panel~(d)) exhibits the same behavior up to $N=8192$: the unfused pipeline is pinned below the 100~GB/s LPDDR5 ceiling ($66$--$87\text{ GB/s}$, matching its $84\text{ GB/s}$ measured copy limit), while fused peak-only filtering sustains $299$--$492\text{ GB/s}$ in shared L2/System Level Cache (SLC) and threadgroup memory. At $N=16384$ ($128\text{ KB}$ per transform), the 8060S retains the transform in its 64~KB LDS using a split exchange, whereas Apple Metal enforces a 32~KB threadgroup memory limit and a 576-thread register budget for this kernel; launching the $1{,}024$-thread workgroup required for 16 samples per thread causes the Metal compiler to spill registers to device memory, reducing flat peak-only throughput at $N=16384$ to $65\text{ GB/s}$. Even at $N=16384$, however, the hierarchical screening chain on the M2 evaluates its coarse stages at $(2048, 8192)$ entirely on-chip without register spilling and refines fewer than $1\%$ of pairs, sustaining $1.0$--$2.7\text{ TB/s}$ of equivalent problem throughput across the full $N \in [1024, 16384]$ range ($3.6$--$12\text{ TB/s}$ on the 8060S).

\section{Discussion and Conclusion}
\label{sec:software}
\label{sec:conclusion}

We have presented two complementary methods for accelerating batched matched filtering: fusing the frequency-domain conjugate product and binned peak search directly into the inverse FFT butterflies with 2D cache tiling, and screening candidate pairs through a hierarchy of short low-frequency transforms whose thresholds are calibrated from the matched-filter noise autocovariance $A \propto \mathcal{F}^{-1}\{|H|^2/S_n\}$.

These algorithms are implemented in the open-source \texttt{matchedfilter} library without external FFT dependencies. CPU kernels are written in C and C++ using Google Highway~\cite{highway} with runtime dispatch across SSE4, AVX2, AVX-512, and ARM NEON instruction sets. GPU kernels are written in Slang~\cite{he2018slang} and compiled to Vulkan SPIR-V, Metal, and CUDA PTX, using atomic stream compaction between screening tiers so subsequent tiers launch workgroups only on surviving candidate pairs. Combining operator fusion with covariance-calibrated screening reduces both memory traffic and transform work, accelerating batched matched filtering on CPUs and GPUs without sacrificing single-precision accuracy or signal recovery beyond a user-specified false-dismissal rate.

Together, the 2D $(n_d \times n_t)$ tiling of the correlation batch and the statistical screening hierarchy may also lend themselves to further acceleration on hardware originally designed for machine-learning workloads. Because coarse screening stages only determine whether a data--template pair warrants refinement while surviving candidates are evaluated by the full-bandwidth \texttt{float32} filter, the early screening tiers do not require full single-precision dynamic range and may be able to operate in reduced-precision formats (\texttt{float16}, \texttt{bfloat16}, or \texttt{fp8}). Combined with 2D batch tiling across data segments and templates, such short reduced-precision stages could potentially leverage the tiled matrix and tensor units of modern GPUs and neural processors while halving or quartering the memory footprint of the low-frequency prefix spectra. More broadly, integrating fused and statistically screened correlation primitives into gravitational-wave, pulsar, and seismic search pipelines offers the potential to run bank-wide analyses on commodity workstations and consumer GPUs that have traditionally required dedicated computing clusters.

\section*{CRediT authorship contribution statement}
\textbf{Alexander H. Nitz:} Conceptualization, Methodology, Software, Writing -- original draft, Writing -- review \& editing. \textbf{Keisi Kacanja:} Validation, Writing -- review \& editing. \textbf{Collin D. Capano:} Conceptualization, Writing -- review \& editing.

\section*{Declaration of competing interest}
The authors declare that they have no known competing financial interests or personal relationships that could have appeared to influence the work reported in this paper.

\section*{Data availability}
No external observational data were used for the research described in this article. The open-source \texttt{matchedfilter} source code, benchmark scripts, and measurement records supporting the findings of this study are openly available under the MIT license at \url{https://github.com/ahnitz/matchedfilter}.

\section*{Acknowledgements}
C.D.C.\ acknowledges support from the National Science Foundation under Grant No.\ PHY-2412341. A.H.N.\ and K.K.\ acknowledge support from the National Science Foundation under Grant No.\ PHY-2309240. The authors thank the developer communities of Google Highway and Shader Slang for their open-source toolchains.

\section*{Declaration of generative AI and AI-assisted technologies in the manuscript preparation process}
During the preparation of this work the authors used Claude (Anthropic) and Gemini (Google) in order to assist with implementing code routines; all algorithmic concepts, mathematical derivations, software architectures, and experimental verifications were directed and validated by the authors. After using these tools, the authors reviewed and edited the content as needed and take full responsibility for the content of the published article.

\bibliographystyle{elsarticle-num}
\bibliography{references}

\end{document}